\documentclass[sigconf]{acmart}

\renewcommand\footnotetextcopyrightpermission[1]{}
\acmConference[SECRYPT 2026]{}{16--18 July, 2026}{Porto, Portugal}
\acmDOI{}
\acmISBN{}
\setcopyright{none}

\AtBeginDocument{%
  \providecommand\BibTeX{{%
    Bib\TeX}}}
\usepackage{tabularx}
\newcolumntype{L}[1]{>{\raggedright\arraybackslash}p{#1}}
\newcolumntype{C}[1]{>{\centering\arraybackslash}p{#1}}
\newcolumntype{R}[1]{>{\raggedleft\arraybackslash}p{#1}}
\def\BibTeX{{\rm B\kern-.05em{\sc i\kern-.025em b}\kern-.08em
    T\kern-.1667em\lower.7ex\hbox{E}\kern-.125emX}}

\usepackage{graphicx}
\usepackage{multirow}
\usepackage{amsmath}
\usepackage{subcaption}
\usepackage{acronym}
\usepackage{orcidlink}
\usepackage{microtype}

\acrodef{gan}[GAN]{\emph{Generative Adversarial Network}}
\acrodef{cavs}[CAVs]{\emph{Connected and Autonomous Vehicles}}

\acrodef{asr}[ASR]{Attack Success Rate}

\newcommand{\al}{\textit{et al}.}

\begin{document}

\title{Assessing the Operational Impact of Poisoning Attacks over Augmented 3D Point Cloud Public Datasets \\ for Connected and Autonomous Vehicles}

\titlenote{This is the authors' accepted version of a paper to appear in the Proceedings of the 23rd International Conference on Security and Cryptography (SECRYPT 2026) - Volume 1, ISBN 978-989-758-858-7, ISSN 2184-7711, pages 712-722, Porto, Portugal, July 16 -- 18, 2026. The final published version will be available through SCITEPRESS.}

\author{Marwan Lazrag \orcidlink{0000-0002-0203-9676}}
\affiliation{%
  \institution{SAMOVAR, Télécom SudParis, Institut Polytechnique de Paris}
  \city{Palaiseau}
  \country{France}
}
\email{marwan.lazrag@telecom-sudparis.eu}

\author{Badis Hammi \orcidlink{0000-0002-4470-6406}}
\affiliation{%
  \institution{SAMOVAR, Télécom SudParis, Institut Polytechnique de Paris}
  \city{Palaiseau}
  \country{France}
}
\email{badis.hammi@telecom-sudparis.eu}

\author{Lorena Gonzalez-Manzano \orcidlink{0000-0002-3490-621X}}
\affiliation{%
  \institution{Universidad Carlos III de Madrid}
  \city{Leganes}
  \country{Spain}
}
\email{lgmanzan@inf.uc3m.es}

\author{Joaquin Garcia-Alfaro \orcidlink{0000-0002-7453-4393}}
\affiliation{%
  \institution{SAMOVAR, Télécom SudParis, Institut Polytechnique de Paris}
  \city{Palaiseau}
  \country{France}
}
\email{joaquin.garcia-alfaro@telecom-sudparis.eu}

\begin{abstract}
Poisoning attacks against public datasets lead to major concerns, such
as (i) misclassification of perceived objects when the poisoned data
is used for training and (ii) embedding of backdoors that may
eventually be triggered later on, when specific conditions in the
system apply over the learned models. Its impact over data
augmentation models is unclear. While data augmentation reduces the
likelihood of poisoning attack success, some valid questions remain.
Is data augmentation affecting the impact of poisoning attacks? can it
increase the number of poisoned samples or injected backdoors? We
explore in this paper some of these questions. We assess the effects
of augmenting poisoned 3D point cloud datasets and validate that
poisoning is able to evade the sanitizing nature of augmentation
techniques when using the concrete case of \ac{gan} techniques to
exemplify the case of data augmentation processing. We also validate
that poisoning propagates over the augmented datasets and perturbs the
decision made by general-purpose classifiers, in the
end. All the experimental material (including tools, datasets, and  classifiers) is publicly available, to facilitate reproducibility and to foster further research in the topic.
\end{abstract}

%\ccsdesc[500]{In-vehicle security~Software vulnerabilities}
%\ccsdesc[300]{Security analysis~SDV ecosystem security}
%\ccsdesc{Do Not Use This Code~Generate the Correct Terms for Your Paper}
%\ccsdesc[100]{Do Not Use This Code~Generate the Correct Terms for Your Paper}

\keywords{Connected and Autonomous Vehicle (CAV), 3D Point Cloud, LiDAR, CCAM, Poisoning Attack, Dataset, Data Augmentation, GAN, Data Sanitization.}

\maketitle

\section{\uppercase{Introduction}}
\label{sec:introduction}

Connected and Autonomous Vehicles (CAVs)\footnote{For the remainder of this paper, the terms ``Connected Autonomous Vehicles (CAVs)" and ``Autonomous Vehicles (AVs)" are used interchangeably in the context of perception-level processing and LiDAR sensor inputs. Distinctions related to connectivity or higher-level vehicular functions are outside the scope of this work.} have progressed rapidly from experimental prototypes to deployed systems, driven by advances in sensing, computation, and connectivity~ \cite{sghaier2025advancing} \cite{A_review_on_autonomous_vehicles_Progress_methods_and_challenges}. Central to CAV operation is perception: a safety-critical subsystem that fuses heterogeneous sensor data, mainly cameras, Light Detection and Ranging (LiDAR), and Radar, to build a reliable, real-time model of the surrounding environment for downstream planning and control \cite{A_State_of_the_Art_Review_on_Attacks_and_Defense_Mechanisms_for_LiDAR_on_Autonomous_Vehicles}. The geometric richness and metric fidelity of LiDAR point clouds, in particular, present both an opportunity and a challenge: they enable precise spatial reasoning (e.g., object localization and shape estimation) but require specialized representations and learning methods to handle sparsity, occlusion, and sensor noise~\cite{Research_advances_and_challenges_of_autonomous_and_connected_ground_vehicles}.

\subsection{Motivation}
The widespread adoption of 3D point cloud deep learning techniques has
significantly advanced the ability of  autonomous vehicles to recognize and
classify objects on the road, enhancing their navigation and
decision-making capabilities~\cite{chen2019cooper}. These techniques
are fundamental for real-time object recognition, scene understanding,
and ultimately, safe autonomous navigation in complex traffic
environments. Despite this progress, a critical gap exists: there is
currently no comprehensive simulator tailored to holistically evaluate
these techniques under diverse conditions.

The validation of such systems generally necessitates comprehensive
simulation environments~\cite{Platelet_Pioneering_Security_and_Privacy_Compliant_Simulation_for_Intelligent_Transportation_Systems_and_V2X}. Among the state-of-the-art tools, Waymo
Simulation
City\footnote{\href{https://waymo.com/blog/2021/07/simulation-city}{https://waymo.com/blog/2021/07/simulation-city}}
stands out as a highly efficient simulator, but it is proprietary,
restricting its accessibility to external researchers. Open-source
alternatives such as
CARLA\footnote{\href{https://github.com/carla-simulator/carla/}{https://github.com/carla-simulator/carla/}}
and Baidu
Apollo\footnote{\href{https://github.com/ApolloAuto/apollo}{https://github.com/ApolloAuto/apollo}}
offer notable advantages, including high customizability, modularity,
and realistic graphics/physics~\cite{yang2021survey}. These simulators
enable researchers to design and test custom scenarios. However, they
rely heavily on scenario creation using existing
datasets~\cite{kaur2021survey}, which are often insufficient in size
and diversity to simulate prolonged and complex driving
scenarios~\cite{li2024choose}.

To address the limitations of dataset size, researchers frequently
employ data augmentation techniques (e.g., Generative Adversarial Networks (GAN)) to augment data and generate synthetic point cloud
samples~\cite{sarmad2019rl}. This approach enables the creation of
longer and more comprehensive simulation scenarios, expanding the
scope of testing and experimentation~\cite{cheng2021dense}. However,
these augmented datasets often rely on publicly available datasets
contributed by other users.

Some studies suggest that data augmentation techniques exhibit a sanitizing effect on data \cite{A_Survey_on_GAN_Techniques_for_Data_Augmentation_to_Address_the_Imbalanced_Data_Issues_in_Credit_Card_Fraud_Detection}\cite{GAN_Based_Data_Augmentation_and_Anonymization_for_Skin_Lesion_Analysis_A_Critical_Review}\cite{Data_Augmentation_Revisited_Rethinking_the_Distribution_Gap_between_Clean_and_Augmented_Data}, as they tend to generate synthetic data that reflects the most common features of the original dataset features, which are, by definition benign.
In this context, \textit{Karra} \al \cite{SanitAIs_Unsupervised_Data_Augmentation_to_Sanitize_Trojaned_Neural_Networks}  exploit unsupervised data augmentation, a self-supervised approach, to mitigate backdoor/Trojan attacks, but their evaluation is restricted to a single manipulation type (Trojan triggers) and to relatively modest poisoning rates (generally 10\% and a maximum of 20\%) \cite{SanitAIs_Unsupervised_Data_Augmentation_to_Sanitize_Trojaned_Neural_Networks}. 
Similarly, \textit{Qin} \al \cite{Learning_the_unlearnable_Adversarial_augmentations_suppress_unlearnable_example_attacks} show that carefully chosen augmentations can suppress the effect of unlearnable or poisoned examples, although their pipeline relies on a verification step prior to augmentation to filter data used for augmentation.
\textit{Rebuffi} \al \cite{Data_augmentation_can_improve_robustness} highlight that augmentation, when coupled with weight averaging, may mitigate robust overfitting and increase adversarial robustness, reinforcing the notion that augmentations encourage learning of high-level stable benign features.
In medical imaging \cite{GAN_Based_Data_Augmentation_and_Anonymization_for_Skin_Lesion_Analysis_A_Critical_Review}, GAN have been used to augment datasets with synthetic images that capture these shared, representative characteristics. 

\subsection{Research gap and hypothesis}

Collectively, these results are encouraging and converge on the claim that augmentation can, under certain conditions, mitigate the influence of data corruption. Nonetheless, such findings are strongly context-dependent and often tested under carefully bounded scenarios, particularly 2D image benchmarks, and thus do not generalize automatically to all data domains. Crucially, they rarely account for the unique challenges posed by 3D point clouds.
The 3D point-cloud modality used for automotive perception differs fundamentally from 2D imagery. Mainly, compared to 2D image domains, point cloud data encode geometric, spatial, and structural continuity, making them especially sensitive to subtle perturbations. Augmentation methods designed to replicate or perturb these structures may inadvertently reinforce adversarial manipulations embedded in the dataset, amplifying their prevalence during model training. Therefore, This divergence highlights a critical research gap.

We believe that this dependency on public datasets, in the case of 3D point-cloud,  creates a critical attack surface.
If a malicious actor introduces a poisoned dataset into this
ecosystem, the augmentation process can exacerbate the poisoning,
resulting in significantly compromised simulation scenarios. As
discussed in the threat model, this could lead to catastrophic
outcomes: the deployed point cloud classifier might fail to accurately
recognize critical 3D objects such as roads, vehicles, or pedestrians.
Furthermore, the adversary could embed a backdoor within the dataset,
triggering a specific behavior from the vehicle under certain
conditions. In autonomous driving scenarios, such failures can result
in severe safety and operational consequences.

\subsection{Contributions of this paper}
Given these risks, it is essential to evaluate the sanitizing effect of data augmentation on 3D point-cloud data used in automotive perception. In this paper we present a focused case study on GAN-based augmentation and investigate whether common augmentation pipelines amplify or attenuate poisoned examples, and how such effects propagate to downstream decision-making in CAVs. We conduct an operational impact assessment, which consists of estimating the impact of interrupting services and functionalities of a system, e.g., inner functionalities or associated processes, due to an attack. The main contributions are listed below:
\begin{itemize}
\item  We present an empirical evaluation study of the sanitization effects of data augmentation on 3D point cloud data for CAVs.  
\item We analyze the impact of poisoned public datasets on downstream classification tasks when these datasets are subjected to augmentation, highlighting potential risks for decision-making in CAVs. 
\item We release the complete implementation, including code and datasets, to facilitate reproducibility and future research on poisoning-aware augmentation strategies.  
\end{itemize}

\medskip

\noindent The remainder of this paper is organized as follows: 
 Section~\ref{sec:rw} surveys related work.
Section~\ref{sec:proposal} presents our proposal and modeling choices
(adversary and impact models). Section~\ref{sec:results} evaluates our
experimental results. Section~\ref{sec:conclusion} concludes the paper.

\section{\uppercase{Related Work}}
\label{sec:rw}

Data augmentation, commonly used to increase the diversity and size of
training datasets, for instance to avoid bias for having lack of
representation of a particular group~\cite{sharma2020data}, has been
applied in many contexts, such as computer vision~\cite{yang2022image}
or natural language processing~\cite{shorten2021text}, among others. A
widespread data augmentation technique is the use of
\ac{gan}~\cite{bissoto2021gan}.

\begin{table}[!h]
  \centering
  \caption{Related work comparison}
    \begin{tabular}{|p{6em}|p{9em}|c|}
    \hline
    \multicolumn{1}{|c|}{Ref.} & \multicolumn{1}{c|}{Poisoning} & \multicolumn{1}{c|}{Operational} \\
        \multicolumn{1}{|c|}{\textbf{}} & \multicolumn{1}{c|}{\textbf{ }} & \multicolumn{1}{c|}{Impact} \\
    \hline
    \cite{li2019pu} & $\checkmark$ point cloud upsampling adversarial attacks & x \\
    \hline
    \cite{koh2023simple} & $\checkmark$ indoor 3D scene synthesis attacks & x \\
    \hline
    \cite{xiang2019generating} & $\checkmark$ 3D adversarial point clouds generation & x \\
    \hline
    \cite{hamdi2020advpc} & $\checkmark$ 3D adversarial perturbations (transferable) & x \\
    \hline
    \cite{wang2024pointapa} & $\checkmark$ availability poisoning attacks in 3D point clouds & x \\
    \hline
    \cite{alsereidi2024data} & $\checkmark$ data poisoning with EEG label-flipping & x \\
    \hline
    \cite{qiancomprehensive2024} & $\checkmark$ data augmentation for 3D adversarial examples & x \\
    \hline
    \cite{hapuarachchi2025securing} & $\checkmark$ error-minimizing attacks & x \\
    \hline
    Ours & $\checkmark$ poisoned 3D point cloud attacks & $\checkmark$ \\
    \hline
    \end{tabular}%
  \label{tab:RW}%
\end{table}%

In the field of 3D point clouds, many approaches have used \ac{gan}
for data augmentation purposes. For instance,~\cite{li2019pu}
introduces PU-GAN, a GAN-based framework designed to upsample sparse
3D point clouds, enhancing data density and uniformity. Another
example is~\cite{koh2023simple}, which proposes an image-to-image
\ac{gan} framework that generates high-resolution RGB-D images from
incomplete point cloud projections.

Attacks against 3D point clouds is also a well studied area.
\cite{wang2024pointapa} presents a novel attack method targeting 3D
point cloud models. The attack disrupts model availability by
embedding class-specific rotations as imperceptible shortcuts into
poisoned point clouds. \cite{xiang2019generating} explores methods to
create adversarial examples by generating imperceptible perturbations
on 3D point cloud data that deceive a deep learning model. Another
example is~\cite{hamdi2020advpc}, which introduces AdvPC, a method for
generating adversarial perturbations on 3D point clouds that are
highly transferable across different models.

Few proposals address the analysis of poisoning attacks under the use of
data augmentation. \cite{alsereidi2024data} focuses on label-flipping
attacks and the use of GAN-generated electroencephalogram data to
compromise model performance in a federated learning environment.
However, the environment, attacks, and data differ from those in this
paper. Complementary to our work, \cite{qiancomprehensive2024} studies
the transferability of 3D adversarial examples by applying different
augmentation techniques, namely drop points, flip, rotation, scale,
shear and translation, and analyzing their effects on different
models. They conclude that data augmentation has a negative impact on
the attack success rate while improving transferability of adversarial
examples. In contrast, we analyze whether data augmentation affects
the operational impact of poisoning attacks. Also, \cite{hapuarachchi2025securing} uses data augmentation against error-minimizing attacks in the context of traffic sign recognition systems in autonomous vehicles.

A summary of analysed works is depicted in Table \ref{tab:RW}, identifying if poisoning is considered and how and if operational impact is somehow addressed. While poisoning has been studied in different ways, its analysis together with operational impact has not been considered so far.

\section{\uppercase{Proposal}}
\label{sec:proposal}

We introduce in this section the threat and  impact models assumed in our work. The former formalizes the knowledge and capabilities assumed from a potential adversary who may perpetrate some poisoning attacks over our motivational scenario. The latter formalizes the impact assessment of the adversary attacks, over the same scenario.

\subsection{Threat Model}
\label{sec:models}

We assume an adversary whose objective is to degrade the victim
model's performance on the clean test distribution, ultimately forcing
it to converge toward random guessing after training on a poisoned 3D
point cloud dataset. We first assume the baseline condition that the model $M$ accurately classifies a clean sample $\mathbf{x}$ as its true label $Y_{\text{true}}$ \cite{kurakin2018adversarial}:
\begin{equation}
    M(\mathbf{x}) = Y_{\text{true}}
\end{equation}

By introducing poisoned data into a data augmentation process, the
adversary aims at ensuring that the newly generated dataset contains
even more poisoned samples. This poisoning is based on generating a perturbed sample $\mathbf{x}'$ from a clean sample $\mathbf{x}$ by injecting a small perturbation $\rho$, such that:
\begin{equation}
    \mathbf{x}' = \mathbf{x} + \rho
\end{equation}

This corrupted data can lead to two major consequences: (i) misclassification of perceived objects when the new dataset is used for training, and (ii) embedding and amplifying a backdoor within the generated dataset, triggering specific behaviors in systems trained on this data. The goal of the attacker is to disrupt the model $M$, leading to the misclassification:
\begin{equation}
    M(\mathbf{x}') = Y' \quad | \quad Y' \neq Y_{\text{true}}
\end{equation}

We assume a clean-label adversary with partial control over the training data. Specifically, the adversary can inject poisoned 3D point cloud samples into the training set at an incremental poisoning rate (e.g., 0\% to 40\%), while preserving the original class labels.\footnote{This clean-label assumption is commonly used in poisoning attack literature~\cite{wang2024pointapa}.} This scenario may occur when an adversary injects malicious samples to public datasets that are later collected and used to train the victim model. However, the adversary is constrained in several ways. They have access only to the training data and no knowledge of external tools, pre-trained models, or surrogate architectures that could facilitate the generation of effective poison instances. In addition, they lack visibility into the victim’s training process, including model architecture, loss function, and hyperparameter choices. Finally, the adversary cannot prevent the defender from manually inspecting the labels of poisoned samples.

\subsection{Impact Model}
\label{subsec:impactModel}

To assess the impact of the attacks perpetrated by the adversary, we assume the concept of operational impact quantification and the use of business logic modeling introduced in~\cite{Lazrag2025}. The first part, on the evaluation of the operational impact quantification, mainly relies on the computation of two functions:
\begin{itemize}
    \item \textbf{Asset criticality evaluation function:} it identifies the criticality of assets within a given system based on how attacks can propagate through and impact those assets. For each asset, this function assigns a criticality value ranging from $0$ to $1$. A value of $0$ indicates that the asset is not impacted by the attack at all, while $1$ indicates that the impact of the attack over such an assets is at its maximum.
    \item \textbf{Impact propagation function:} it quantifies how an external event (e.g., a failure or an attack) propagates its effects over the operational functions and the operational processes associated to a given system, in probabilistic terms.
\end{itemize}

The two previous functions assume the existence of a \emph{resource dependency graph}, representing the dependencies between all the assets in the system, and a \emph{mission dependency graph} describing the relationships between system-level functions and operational processes associated with each asset. In both cases, nodes represent assets, system functions, and operational processes, while edges capture the interdependencies between them.

\begin{figure}
        \centering
        \includegraphics[width=\columnwidth]{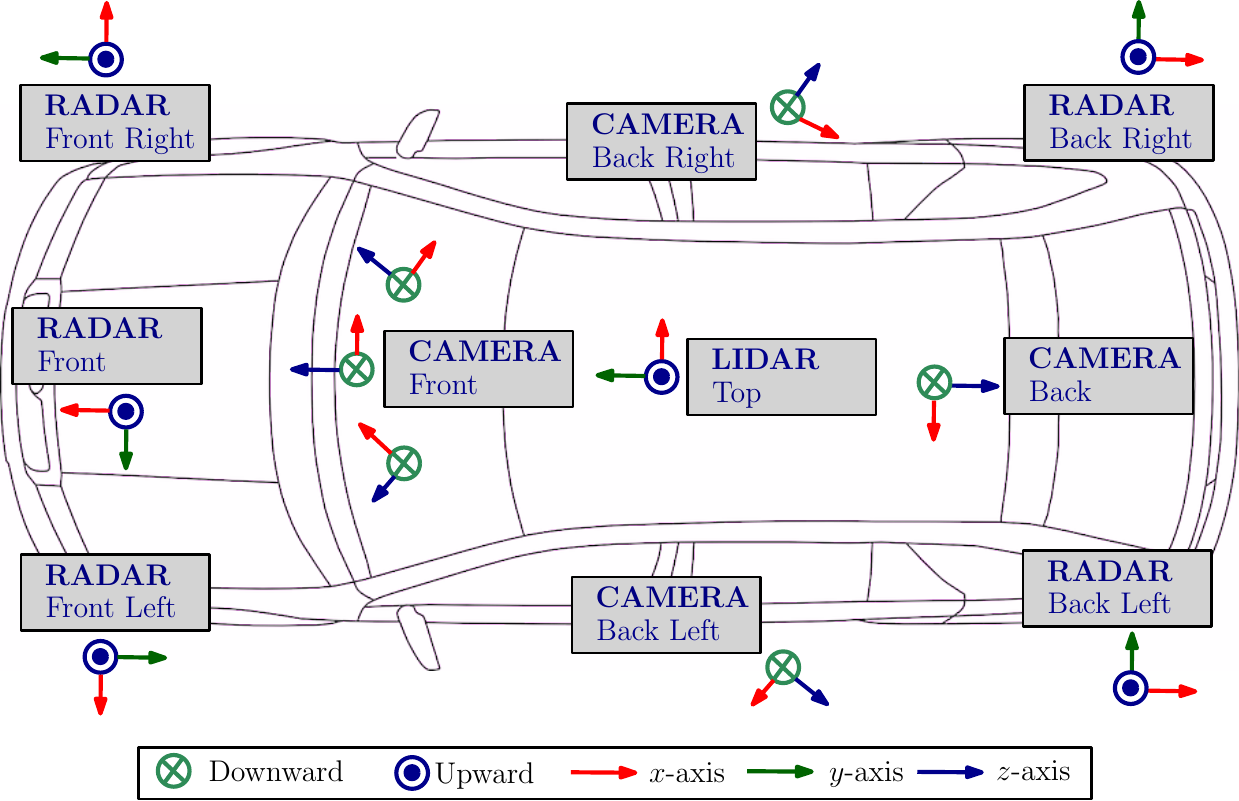}
        \caption{CAV scenario inspired from the nuScenes data collection platform \cite{nuscenes}.}
        \label{fig:BLM1}
\end{figure}

\begin{figure}
        \centering
        \includegraphics[width=\columnwidth]{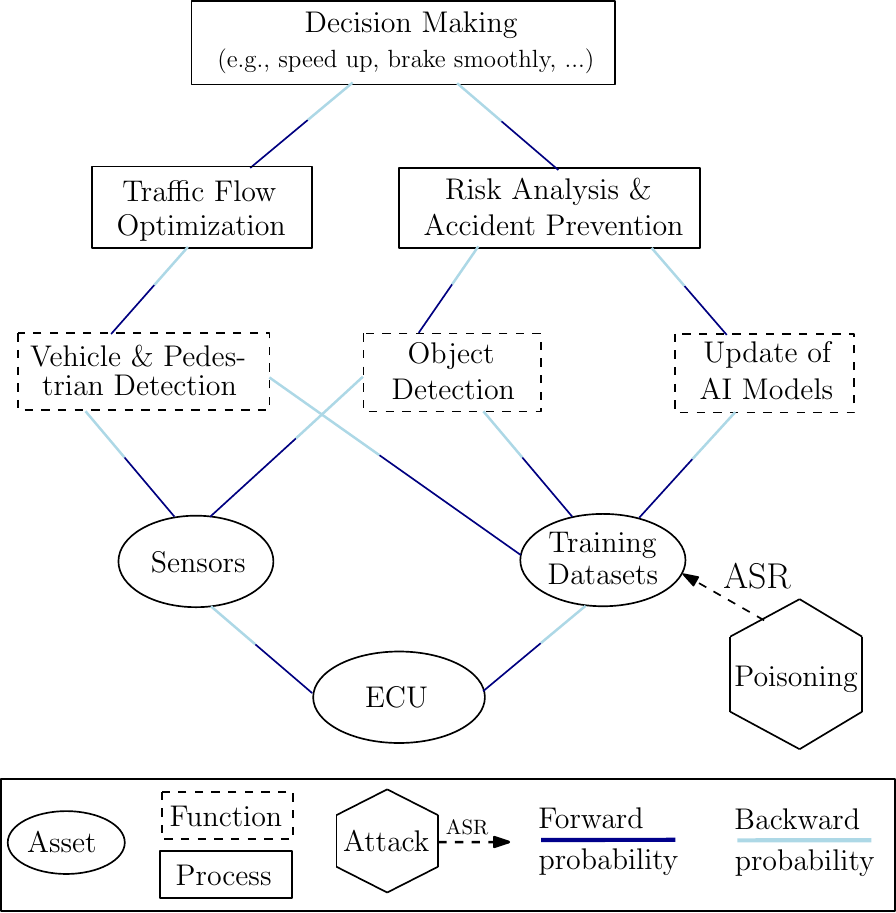}
\caption{Assessing the impact of a poisoning attack against the training datasets of the CAV scenario in Figure~\ref{fig:BLM1}, using the impact propagation model defined in~\cite{Lazrag2025}.}
\label{fig:BLM2}
\end{figure}

Figure \ref{fig:BLM1} illustrates the system model of a CAV scenario, inspired from the nuScenes dataset\footnote{The nuScenes dataset is a large-scale, multimodal benchmark (camera, LiDAR, radar, IMU and rich 3D annotations) that is widely used in the literature for perception, tracking and prediction tasks. Its standardized evaluation and provided devkit have made it a de facto reference in autonomous-driving research.} \cite{nuscenes}, which details representative assets such as sensors and controllers. Based on this model, we construct the corresponding resource and mission dependency graphs. Figure \ref{fig:BLM2} presents our impact assessment model, combining both types of dependencies into a unified graph\footnote{The tool used to develop the impact-assessment model and to evaluate the effects of poisoning attacks on vehicle operational functions is derived from existing work~\cite{Lazrag2025}. For reproducibility purposes, we release the full implementation and  modifications of the original work and tools as a Docker-deployable codebase (source code, Dockerfiles, and detailed implementation and deployment instructions) at \url{https://github.com/Marwanlz/Assessing_Operational_Impact_Poisoning_3d_PointCloud-CAV}.}. To clearly distinguish between the two subgraphs, assets are depicted as circle-shaped nodes (representing the resource dependency graph), while system functions and operational processes are shown as rectangle-shaped nodes (representing the mission dependency graph).

%%%%%%%%%%%%%%%%%%%%%%%%%%%%%%%%%%%%%%%%%%%%%%%%%%%%%%%%%%%%%%%%%%%%%%%%%%%%%%%%%%%%%%%%%%%%%%%%%%%%%%%%%%%%

%%%%%%%%%%%%%%%%%%%%%%%%%%%%%%%%%%%%%%%%%%%%%%%%%%%%%%

%%%%%%%%%%%%%%%%%%%%%%%%%%%

The example depicted in Figure \ref{fig:BLM2} also describes the correlation between assets, system functions, and operational processes in a representative CAV scenario from \cite{nuscenes}. It contains three assets\footnote{For clarity, all vertices corresponding to non-impacted assets (e.g., Camera) and their directly linked operational functions (e.g., Weather Conditions Analysis, Itinerary Optimization, and so on) have been omitted from the diagram.}: \emph{sensors}, \emph{training datasets} and \emph{Electronic Control Unit (ECU)}; three system functions: \emph{Vehicle \& Pedestrian Detection}, \emph{Object Detection}, and \emph{Update of AI models}; and three operational  processes: \emph{Traffic Flow Optimization}, \emph{Risk Analysis \& Accident Prevention}, and \emph{Decision Making}. The \emph{Decision Making} process represents the adversary’s ultimate target. By launching a poisoning attack against the asset \emph{Training Datasets}, the adversary seeks to disrupt the outcomes of this process, for instance, by misleading the autonomous vehicle and causing it to make incorrect decisions.

In this propagation model, edges connecting vertices represent interdependency probabilities, specifically Forward and Backward probabilities that quantify how degradation propagates between system components. The Attack entity is characterized by an Attack Success Rate (ASR), which measures the initial detrimental impact of the poisoning attack on the targeted operational process. This ASR serves as the initial probability that cascades through the interconnected graph structure, influencing both Forward and Backward probabilities of all connected vertices. The combined effect of these probability interactions enables the quantification of the operational impact stemming from the attack against the root operational process. The ASR thus acts as the propagation catalyst, where the initial attack probability diminishes or amplifies as it traverses the dependency network, ultimately determining the extent of system degradation. A practical demonstration of this impact propagation mechanism is illustrated in Section \ref{subsubsec:ImpactQuantification}, Figure \ref{fig:impact_assessment}.

\section{\uppercase{Experimental methodology}}
\label{sec:results}

In this section, we present our experimental framework and the results obtained. 

%%%%%%%%%%%%%%%%%%%%%%%%%%%%%%%%%%%%%%%%%%%%%%%%%%%%%%%%%%%%%%%%%%%%%%%%%%%%%%%%%%%%%%%%%%%%%%%%%%%%%%%%%%%%%%%%%%%%%%%%%%%%%%%%%%%%%%%%%%%%%%%%%%%%%%%%%%%%%%%%%%%%%%%%%%%%%%%%%%%%%%%%%%%%%%%%%%%%%%%%%%%%%%%%%%%%%%%%%%%%%%%%%%%%%%%%%%%%%%
\subsection{Overview}
\label{subsec:methodology}

%%%%%%%%%%%%%%%%%%%%%%%%%%%%%%%%%%%%%%%%%%%%%%%%%%%%%%%%%%%%%%%%%%%%%%%%%%%%%%%%%
\begin{figure*}[!t]
\includegraphics[width=\linewidth]{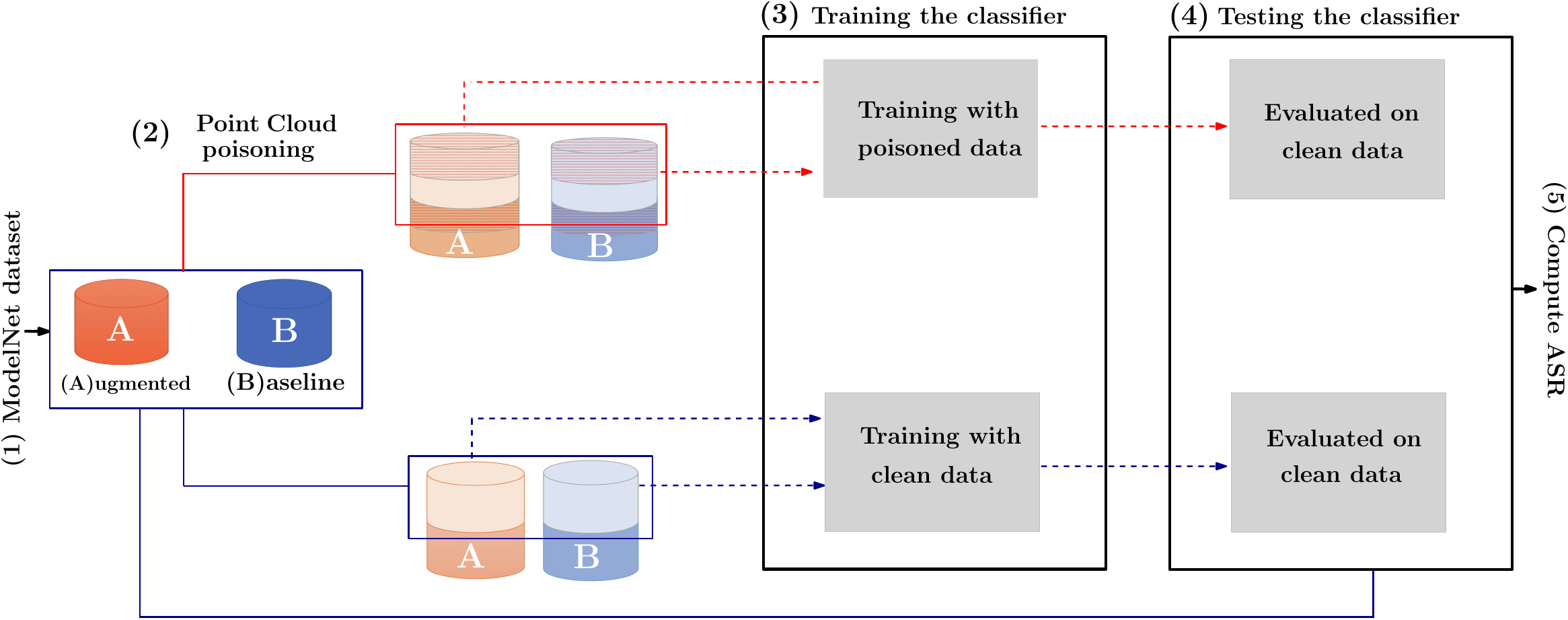}
\caption{Methodology to compute the \ac{asr} associated to the motivational scenario depicted in Figure~\ref{fig:BLM2}. \label{fig:expOverview}}
\end{figure*}
%%%%%%%%%%%%%%%%%%%%%%%%%%%%%%%%%%%%%%%%%%%%%%%%%%%%%%%%%%%%%%%%%%%%%%%%%%%%%%%%%

As Figure \ref{fig:expOverview} shows, we designed two experimental scenarios to evaluate the impact of data poisoning under different conditions:
\begin{enumerate}
\item The classifier is trained on the original (baseline) dataset.
\item The classifier is trained on an augmented dataset, which includes several new samples generated using the 3D-GAN framework \cite{wu2016learning}.
\end{enumerate}

The rationale behind these two scenarios is to assess the effect of poisoning both with and without the use of augmentation techniques. By doing so, we can achieve an operational impact assessment, which consists of estimating the impact of interrupting services and functionalities of a system, e.g., inner functionalities or associated processes, due to the poisoning attack. In both settings, the training datasets are of equal size to ensure a fair and consistent comparison. 
The experiments in each scenario are conducted as follows. 
First, we preprocess the ModelNet dataset\footnote{ModelNet was introduced as a large-scale 3D Computer-Aided Design (CAD) model dataset and is widely used as a benchmark for 3D shape / point-cloud research. It is frequently employed in studies of adversarial attacks and defenses on 3D point clouds.} \cite{wu20153d}, which contains a variety of 3D object models: we select two classes for binary classification (a primary target class and a secondary, non‐target class), split the data into training and test sets, and reserve only the primary‐class training samples for manipulation (all secondary‐class and test samples remain unaltered). 
For training we used 3,000 files (1,500 per class), and for testing we used 600 files (300 per class). The dataset is class-balanced with an 83\%/17\% train/test split. We report exact counts to ensure experimental reproducibility and to allow fair assessment of sampling variability.

Next, in both scenarios, we train our classifier on the clean training data, using original 3D objects for the first scenario (baseline scenario) and synthetic 3D objects generated by the 3D-GAN framework for the second (augmented scenario), and evaluate its performance on the untouched test set to establish baseline metrics. Finally, we simulate an adversary by modifying the 3D shapes of a subset of primary‐class samples and injecting these poisoned examples into the training set at incrementally increasing rates (0 \% to 40 \%). For each Poisoning rate, we retrain the classifier on the modified dataset and evaluate it on the same clean test set. This procedure allows us to assess how varying levels of data poisoning, both with and without augmentation, affect the classification accuracy and attack success. 

In contrast to saliency-based point-dropping techniques such as proposed by \textit{Zheng} \al  \cite{Pointcloud_saliency_maps} or point-detach strategies that iteratively remove high-importance points \cite{Adversarial_attack_and_defense_on_point_sets}, we employed a simpler poisoning method, typically, we randomly removed 50\% of the points from each poisoned file, without applying any ranking or importance criterion. 

\begin{figure}[!h]
\begin{center}
\subfloat[]{\includegraphics[width = 0.15\textwidth]{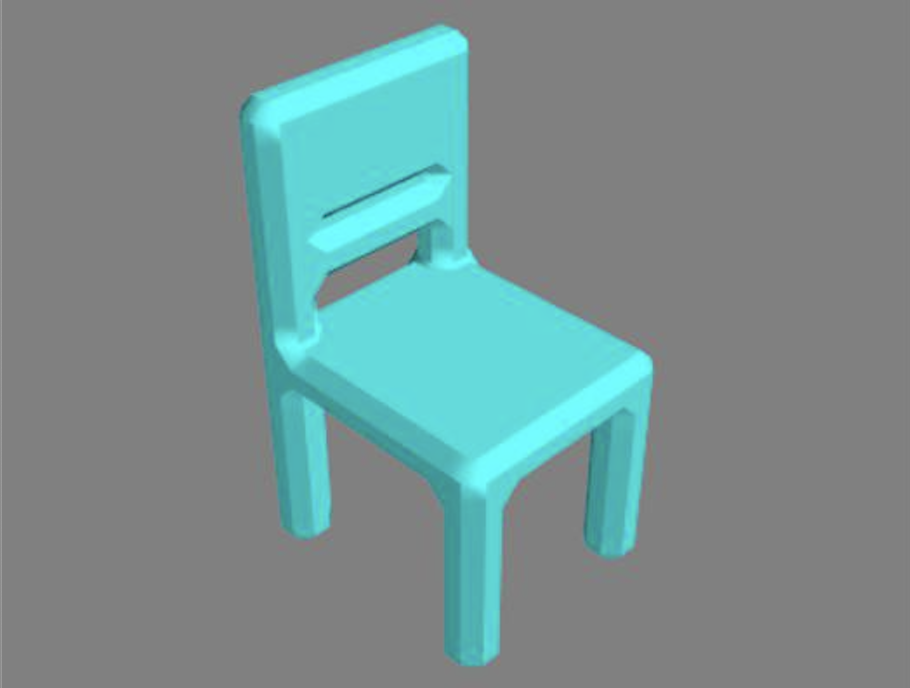}}
\hspace{0.01 cm}
\subfloat[]{\includegraphics[width = 0.15\textwidth]{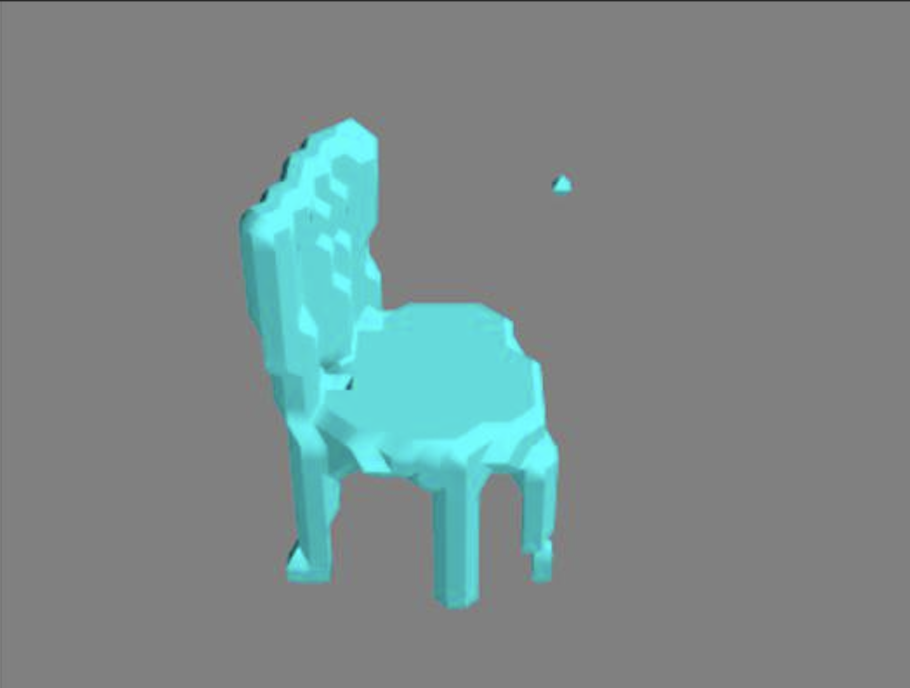}}
\hspace{0.01 cm}
\subfloat[]{\includegraphics[width = 0.15\textwidth]{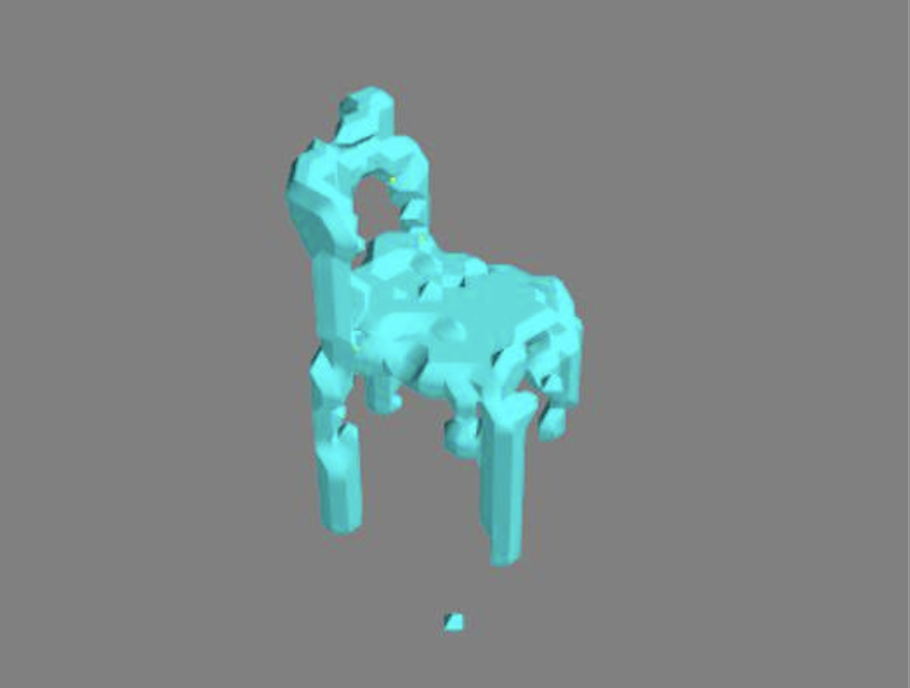}}
\caption{(a) Original object; (b) GAN output from clean data; (c) GAN output from poisoned data.}
\label{fig:methodo}
\end{center}
\end{figure}

It is worth noting that the experiments conducted in the second scenario are preceded by the creation of an augmented dataset using the 3D-GAN framework. In this setting, the poisoning data is deliberately introduced into the training dataset of the 3D-GAN. The newly generated 3D samples are then used to train the classification model for the primary class, whereas the data for the secondary class remains unaltered. 
It is worth noting that, the test set used in this scenario is identical to that of Scenario 1, ensuring a consistent evaluation protocol. Figure \ref{fig:methodo} shows (a) an original 3D object, (b) a synthetic 3D object generated by a GAN trained on the clean dataset, and (c) a synthetic 3D object generated by a GAN trained on a poisoned dataset. This illustration highlights the effects of dataset poisoning on the quality and characteristics of GAN-generated 3D objects.
%%%%%%%%%%%%%%%%%%%%%%%%%%%%%%%%%%%%%%%%%%%%%%%%%%%%%%%%%%%%%%%%%%%
\subsection{Experimental setup and metrics}
\label{subsec:Setup}

The experimental setup is designed to evaluate the classification performance against data poisoning attacks\footnote{The experiments (including source code, original datasets, augmented datasets, poisoned dataset, data augmentation code, and data classification code) are publicly available at \url{https://github.com/Marwanlz/Assessing_Operational_Impact_Poisoning_3d_PointCloud-CAV}}.
The experiments are conducted on a machine equipped with an Intel i7-11850H CPU processor and an NVIDIA RTX A4000 GPU. Data augmentation via 3D-GAN is conducted using PyTorch\footnote{\url{https://pytorch.org}}. Training of the binary classifier is conducted using TensorFlow\footnote{\url{https://tensorflow.org}}. 
The classifier is built using an existing InceptionNet architecture, which was adapted by us for binary classification.

The classifier is trained with the Adam optimizer \cite{kingma2014adam}, using binary cross-entropy as the loss and a sigmoid activation on the output to produce probabilistic predictions. The Adam optimizer provides adaptive learning-rate updates that speed and stabilize convergence. Binary cross-entropy is the canonical log-loss for two-class problems. And the sigmoid yields well-interpretable posterior scores for thresholding or calibration. These settings are standard and widely adopted for binary classification in deep-learning work \cite{Binary_cross_entropy_with_deep_learning_technique_for_image_classification}\cite{Deep_learning_with_Python}\cite{Deep_learning_based_vulnerability_detection_in_binary_executables}.
The training of the classifier is conducted over $20$ epochs with a batch size of $32$, which provides a good trade-off between convergence speed and performance consistency on the validation dataset. The model performance is monitored at each epoch, using validation accuracy to ensure stable learning.

To align with the literature, we evaluate the classifier's robustness using the F1 score. However, We  also use the Matthews correlation coefficient (MCC) metric. F1 is a harmonic mean of precision and recall, it compactly measures a detector’s ability to find true positives while limiting false alarms, making it well suited for tasks that prioritise the positive class. Whereas MCC is a correlation coefficient that uses all four confusion-matrix entries and ranges from $-1$ (total disagreement) to $+1$  (perfect prediction) (with $0$ means random prediction). Because it accounts for true negatives as well as positives, MCC provides a balanced, prevalence-insensitive assessment of overall classifier quality. 
Reporting both metrics is essential, F1 reflects the precision–recall trade-off for the target class, while MCC reveals overall performance and exposes pathological behaviour on imbalanced data (e.g., models that attain high F1 by exploiting a tiny positive class but fail on negatives). Together they prevent over-claiming detector performance and make hidden failure modes visible.

We also report a third metric to quantify poisoning effectiveness on classifier outputs: the Attack Success Rate (ASR), Formally:
    $$
    \text{ASR} = \frac{\text{FN}_{\text{after attack}} - \text{FN}_{\text{before attack}}}{\text{Total number of samples}} \cdot 100
    $$
where $N$ is the total number of tested samples, and $FN_{\text{before attack}}$ $FN_{\text{after attack}}$ are the counts of false negatives before and after the poisoning, respectively.
This ASR measures the increase in missed detections attributable to the attack, normalized by the evaluation set size. 
We focus on false negatives because poisoning in our threat model aims to increase wrongful non-detections (samples that should be flagged but are not).  In other words, we consider only false negatives (not false positives) because the evaluation targets the primary class exclusively. In this setting, true positives are primary-class samples correctly identified; false negatives are primary-class samples incorrectly classified due to the poisoning; and false positives are non-primary samples incorrectly labeled as primary. This focus ensures that reported metrics capture degradations in recognition of the targeted class.   

All experiments were repeated five times. The reported values are the mean across runs. Standard deviations were consistently small, demonstrating the stability of the results. 
For clarity, in the next section, we omit the standard deviations from the main figures and tables.

%%%%%%%%%%%%%%%%%%%%%%%%%%%%%%%%%%%%%%%%%%%%%%%%%%%%%%%%%%%%%%%%%%%%%%%%%%%%%%%%%%%%%%%%%%%%%%%%%%%%%%%%%%%%%%%%%%%%%%%%%%%%%%%%%%%%%%%%%%%%%%%%%%%%%%%%%%%%%%%%%%%%%%%%%%%%%%%%%%%%%%%%%%%%%%%%%%%%%%%%%%%%%%%%%%%%%%%%%%%%%%%%%%%%%%%%%%%%%%
%%%%%%%%%%%%%%%%%%%%%%%%%%%%%%%%%%%%%%%%%%%%%%%%%%%%%%%%%%%%%%%%%%%

%%%%%%%%%%%%%%%%%%%%%%%%%%%%%%%%%%%%%%%%%%%%%%%%%%%%%%%%%%%%%%%%%%%

\subsection{Experimental results}
\label{subsec:ExperimentalResults}

In the following, the analysis of the poisoning attack is presented (Section \ref{subsubsec:Metrics}), together with the impact quantification study (Section \ref{subsubsec:ImpactQuantification}).

\subsubsection{Poisoning analysis} 
\label{subsubsec:Metrics}

\begin{figure*}[ht!]
\includegraphics[width=\linewidth]{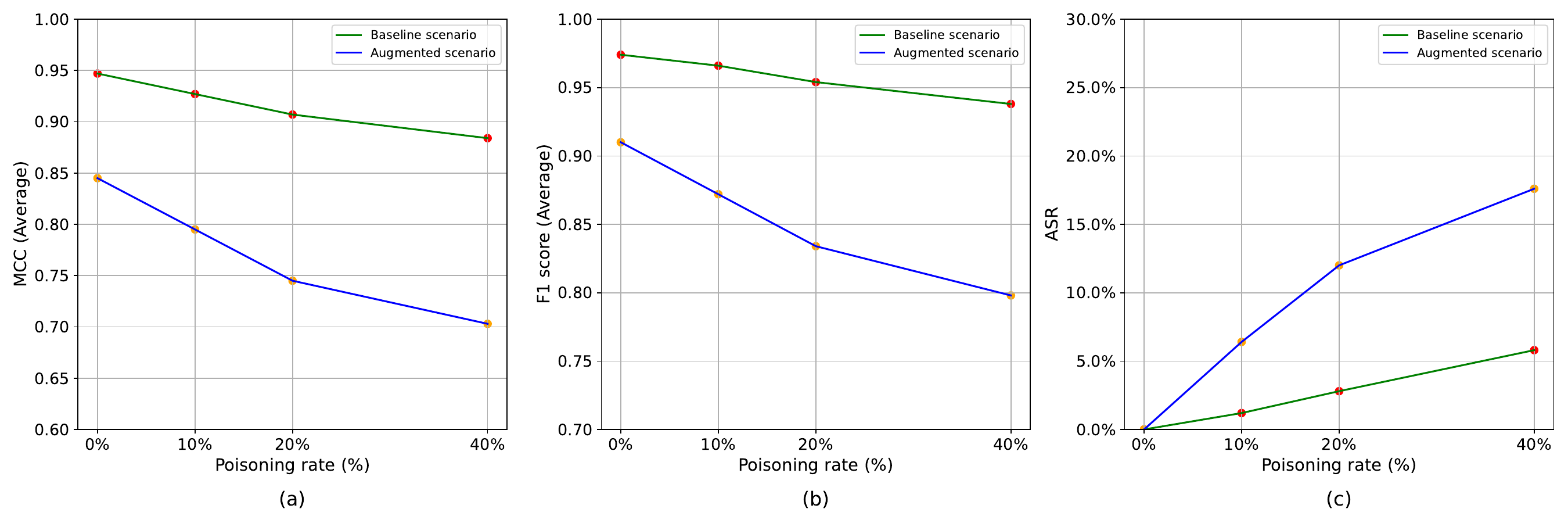}
\caption{Dependencies between metrics and poisoning rate: (a) Dependency between MCC and Poisoning rate; (b) Dependency between F1 score and Poisoning rate; (c) Dependency between ASR and Poisoning rate.}
\label{fig:classification_Results}
\end{figure*} 

\begin{figure*}[!h]
\begin{center}

\includegraphics[width =\textwidth]{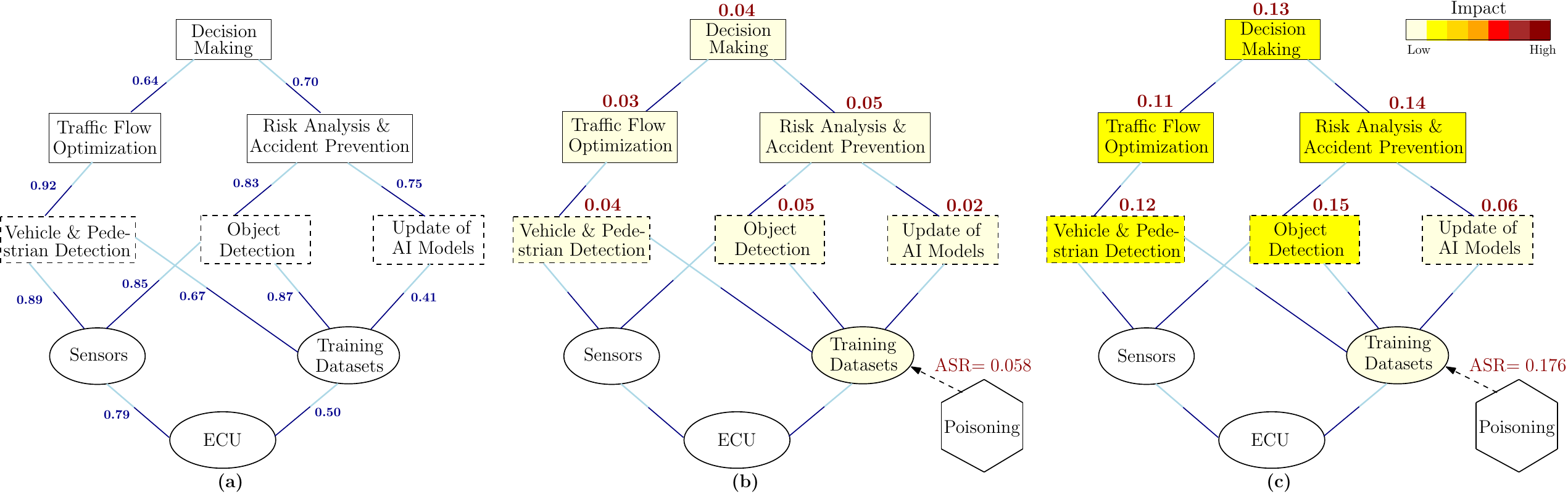}
\caption{Impact assessment model: (a) Nominal system state; (b) Baseline scenario with a 40\% poisoning rate; (c) Augmented scenario with a 40\% poisoning rate.}
\label{fig:impact_assessment}
\end{center}
\end{figure*} 

The classification metrics results are summarised in Figure \ref{fig:classification_Results} for both the Baseline scenario and the Augmented scenario (see Section \ref{subsec:methodology}), using the metrics defined in Section \ref{subsec:Setup}.
In both scenarios, only the training set of the primary class was manipulated by injecting poisoned samples at rates ranging from $0\%$ (without poisoning) to $40\%$. 
The secondary class and the entire test set were left unchanged to isolate the effect of poisoning on model performance.

In each figure, the x-axis displays the reported classification metric (MCC, F1, ASR) and the y-axis indicates the poisoning rate (in \%).
Across both evaluation scenarios (Baseline and Augmented), the classifier's performance degrades monotonically with increasing poisoning: MCC and F1 decline while Attack Success Rate (ASR) rises, and the performance gap between scenarios widens as the poisoning load grows. At low poisoning levels the two setups differ only marginally ($\approx$ 0.1 point in MCC and a similar delta in F1), indicating comparable initial robustness.
By 20\% poisoning the Baseline still yields strong detection (MCC $\approx$ 0.90, F1 $\approx$ 0.95) with a small ASR (2.8\%). At the highest injection rates the divergence becomes pronounced: the Baseline records MCC = 0.88, F1 = 0.93, ASR = 5.8\%, whereas the Augmented setup drops to MCC = 0.70, F1 = 0.79, ASR = 17.6\%.

These results indicate that, under our threat model and augmentation pipeline, the augmentation amplifies the impact of poisoned samples: rather than diluting adversarial artefacts, the augmentation process appears to reinforce distributional modes that the attacker exploits, increasing misclassification rates and overall attack effectiveness. This behavior is consistent with a mechanism where augmentation shifts model emphasis toward features present in both clean and poisoned examples, thus enlarging the adversary’s effective feature space. We recall that all reported values are averages over repeated runs.

%%%%%%%%%%%%%%%%%%%%%%%%%%%%%%%%%%%%%%%%%%%%%%%%%%%%%%%%%%%%%%%%%%%%%%%%%%%%%%%%%%%%%%%%%%%%%%%%%%%%%%%%%%%%%%%%%%%%%%%%%%%%%%%%%%%%%%%%%%%%%%%%%%%%%%%%%%%%%%%%%%%%%%%%%%%%%%%%%%%%%%%%%%%%%%%%%%%%%%%%%%%%%%%%%%%%%%%%%%%%%%%%%%%%%%%%%%%%%%
\subsubsection{Impact Quantification}
\label{subsubsec:ImpactQuantification}

%%%%%%%%%%%%%%%%%%%%%%%%%%%%%%%%%%%%%%%%%%%%%%
\begin{table}[!h]
\centering
\caption{Assessment of the impact on the operational function \textit{Decision Making} based on ASR values}
\label{tab:OperationalImpact}
\centering
\begin{tabular}{|p{1.5cm}||p{1.5cm}|p{1.5cm}|p{1.5cm}|}
\hline
Scenario & Poisoning Rate & ASR & Operational Impact \\
\hline
\hline
\multirow{3}{*}{Baseline} & 10\% & 1.2 \% & 1 \% \\
\cline{2-4} & 20\% & 2.8 \% & 2 \% \\
\cline{2-4} & 40\% & 5.8 \% & 4 \% \\
\hline
\hline
\multirow{3}{*}{Augmented } & 10\% & 6.4 \% & 5 \% \\
\cline{2-4} & 20\% & 12 \% & 9 \% \\
\cline{2-4} & 40\% & 17.6 \% & 13 \% \\
\hline
\end{tabular}
\end{table}

%%%%%%%%%%%%%%%%%%%%%%%%%%%%%%%%%%%%%%%%%%%%%%
We use the ASR as a concrete, operational indicator of poisoning effectiveness and feed it into the impact-propagation model from Section \ref{subsec:impactModel} to quantify downstream effects on the \textit{Decision Making} operational function. Concretely, ASR values measured for each poisoning rate are treated as probabilistic inputs to the propagation function, which maps classifier degradation to the likelihood of impaired decision outcomes in the motivational scenarios.

Figure \ref{fig:impact_assessment} illustrates the impact assessment model across three use cases. For clarity, the assets: \emph{Electronic Control Unit (ECU)}, \emph{sensors}, and \emph{training datasets} are replaced by \emph{Asset 1}, \emph{Asset 2}, and \emph{Asset 3}; the system functions: \emph{Vehicle \& Pedestrian Detection}, \emph{Object Detection}, and \emph{Update of AI models} by \emph{Function 1}, \emph{Function 2}, and \emph{Function 3}; and the operational  processes: \emph{Traffic Flow Optimization}, \emph{Risk Analysis \& Accident Prevention}, and \emph{Decision Making} by \emph{process 1}, \emph{process 2}, and \emph{process 3}.
Figure \ref{fig:impact_assessment}.a depicts the nominal (pre-attack) system state where edges encode interdependencies between assets. While the model supports bidirectional impacts, this particular scenario exhibits only downstream propagation, meaning all backward impact probabilities are set to zero. For visual clarity, only the forward impact probabilities (shown in dark blue) are displayed, as the reverse probabilities would contribute no meaningful information to the analysis. %For example, the impact probability from LiDAR sensors to the ECU is 0.79, while the reverse probability is 0.0.} 
In Figure \ref{fig:impact_assessment}.a, the edge values between operational functions and processes (represented as rectangular nodes) were manually assigned based on domain expertise, as modeling these relationships requires in-depth knowledge of vehicle activities~\cite{Lazrag2025}. For interdependencies between assets (represented as circular nodes), edge values were derived from the nuScenes dataset \cite{nuscenes}.

Figure \ref{fig:impact_assessment}.b depicts the Baseline scenario with a $40$\% poisoning rate, while Figure \ref{fig:impact_assessment}.c illustrates the Augmented scenario under the same poisoning rate. These figures demonstrate how an attack propagates its impact across the graph, with interdependencies recalculated accordingly. The degradation level of each asset or operational function (represented as rectangles) is indicated by a red numerical value above each vertex. A color gradient from white (no impact) to yellow (low impact) to dark red (severe impact) visually encodes the extent of degradation on these vertices. This value quantifies the perturbation in the execution and performance of vehicular functions. It ranges from 0 (no impact) to 1 (very high impact), corresponding to 0\% and 100\%, respectively. For example, in the baseline scenario, the operational impact of the poisoning attack on the \emph{Traffic Flow Optimization} function is 0.03 (3\%), as shown in Figure \ref{fig:impact_assessment}.b. In the augmented scenario, the operational impact on the \emph{Traffic Flow Optimization} function reaches 0.11 (11\%), as shown in Figure \ref{fig:impact_assessment}.c.

Table \ref{tab:OperationalImpact} summarizes the impact probability of the poisoning attack on operational functions for both scenarios.
The resulting assessment exhibits two robust patterns. First, impact probability grows monotonically with ASR, denoted . That is, higher attack-induced misclassification directly increases the chance that decision-making functions receive corrupted inputs and produce incorrect outcomes. 
Second, the Augmented scenario consistently yields substantially larger operational impacts than the Baseline for the same poisoning rate. For instance, at 40\% poisoning the assessed probability that \textit{Decision Making} is affected rises from 4\% (Baseline) to 13\% (Augmented), hence, an over threefold increase in impact under our modeling assumptions. 
Hence, the analysis demonstrates a clear and actionable insight: augmentation,  amplifies attack effectiveness and materially increases the probability of operational disruption in decision-making functions.

Finally, while this evaluation focuses on a specific poisoning attack and a GAN-based augmentation technique, the observed increase in misclassification and operational impact is not limited to this setting. Comparable effects may also arise with other attacks, potentially leading to similar performance degradation and increased operational impact under our assessment model. Similarly, augmentation techniques other than GANs that expand the training dataset may influence the propagation of poisoned samples, resulting in comparable amplification effects.

\section{\uppercase{Conclusion}}
\label{sec:conclusion}

We addressed in this paper the issue of poisoning attacks against public datasets. More precisely, the case of poisoning attacks over augmented 3D point cloud public datasets for Connected and Autonomous Vehicles (CAV) scenarios. Poisoning attacks are known to lead to misclassification of perceived objects. Even worse, they can assist adversaries to embed backdoors that may eventually be triggered later on, when specific conditions in the system apply over the learned models. We assessed the operational impact of this attacks over data
augmentation models. While data augmentation reduces the
likelihood of poisoning attack success, we addressed whether data augmentation keeps affecting the impact of poisoning attacks over general purpose classifiers. We experimentally validate that data augmentation can even increase the number of poisoned samples, hence augmenting as well the effects of augmenting poisoned 3D point cloud datasets. We validated that
poisoning is able to evade the sanitizing nature of augmentation
techniques under the concrete case of \ac{gan} techniques. Our results  validate as well that poisoning propagates over the augmented datasets and perturbs the decision made by general-purpose classifiers. Extending this evaluation to other 3D point cloud datasets remains challenging, given the limited availability of such datasets for CAV scenarios.

\section*{\uppercase{Acknowledgements}} 

The work presented in this paper was conducted within the framework of the Horizon Europe AI4CCAM project (grant agreement 101076911), addressing trustworthiness of artificial intelligence in the context of Connected, Collaborative and Automated Mobility. Lorena González and Joaquin Garcia-Alfaro  are being partially supported by Project PID2023-150310OB-I00 (MORE4AIO) funded by MCIU/ AEI / 10.13039/501100011033 / FEDER, UE. We thank the anonymous reviewers for their valuable comments and helpful suggestions.

\bibliographystyle{IEEEtran}

\bibliography{example}

\end{document}